\documentclass[aps,pra,twocolumn,groupedaddress]{revtex4}
\usepackage{graphicx}

\begin{document}


\title{Continuous Dynamical Decoupling with Bounded Controls}


\author{Pochung Chen}
\affiliation{Department of Physics, National Tsing-Hua University,
  Hsinchu 100, Taiwan}


\date{\today}

\begin{abstract}

  We develop a theory of continuous decoupling with bounded controls 
  from a geometric perspective.
  Continuous
  decoupling with bounded controls can accomplish the same decoupling
  effect as the bang-bang control while using realistic control resources 
  and it is robust against systematic implementation errors.  
  We show that the decoupling
  condition within this framework is equivalent to average out error
  vectors whose trajectories are determined by the control
  Hamiltonian. The decoupling pulses can be intuitively designed once
  the structure function of the corresponding $SU(n)$ is known and is
  represented from the geometric perspective.  Several examples are
  given to illustrate the basic idea. From the physical implementation
  point of view we argue that the efficiency of the decoupling is
  determined not by the order of the decoupling group but by the
  minimal time required to finish a decoupling cycle.

\end{abstract}


\maketitle

\def\Bid{{\mathchoice {\rm {1\mskip-4.5mu l}} {\rm
{1\mskip-4.5mu l}} {\rm {1\mskip-3.8mu l}} {\rm {1\mskip-4.3mu l}}}}

\section{Introduction}

Quantum decoherence control has been studied intensively in recent
years due to the fact that it represents one of the main obstacles in
implementing quantum computation. However the problem of decoherence
remains daunting. Although there have been rapid advances in physical
realization of quantum operations on few levels system in which a
single qubit or few qubit can be embedded .  The fidelity for those
quantum operations, such as Rabi rotation, is typically far below the
threshold to make the system scalable.  To increase the fidelity, many
strategies are proposed to counteract the undesired effects resulting
from decoherence. Those strategies can be roughly divided into two
categories, depended on if an encoding scheme is used. Some prominent
examples in which encoding is needed are the quantum error-correction
codes (QECC)\cite{shor95,steane96,knill97} and the decoherence-free
subspace (DFS).\cite{zanardi97,duan97,lidar98} One of the main
drawback of the encoding strategies is the large amount of ancillary
space needed, especially when the qubit is still a rare resource. As a
result, strategies which do not need ancillary systems is highly
desirable.  Closed-loop decoherence control (such as quantum feedback
control \cite{WM92,DJ99,ahn:042301}) and open-loop decoherence control
(such as bang-bang or dynamical decoupling
\cite{LVEK99,LVSL99,PZ99,LVEK00}) fall into this category.  The
open-loop decoherence control typically requires only a limited, {\it
  a priori}, knowledge of the system-environment dynamics.  Empirical
determination of control parameters via quantum tomography has also
been proposed recently.\cite{byrd:012324} In this work we will focus
on the dynamical decoupling, but it is acknowledged that no single
strategy can efficiently eliminated the problem of decoherence at all
levels. Combining or concatenating different strategies is usually
necessary for any real physical system.

In the original bang-bang decoupling framework, arbitrarily strong and
instantaneous control pulses are utilized to induce frequent unitary
interruptions during the evolution of the system.  The control
Hamiltonian is independent of the system dynamics and are judicially
designed to realize an effective decoupling between the system and the
environment. The stringent requirement on the control Hamiltonian
represents a drawback of the bang-bang decoupling.  Experimentally it
is impossible to implement arbitrarily strong and instantaneous pulses
in real physical systems.  An ultra strong and fast control pulse will
also inevitably induce transitions to higher energy levels which are
usually neglected in the analysis of the bang-bang decoupling.
Theoretically is difficulty to describe the evolution with and without
control terms simultaneously.  This also make it difficult to estimate
the robustness of the control and to estimate the error induced by
operational imperfections.  The highly abstract group theory which is
frequently used in the decoupling analysis also make it less intuitive
and difficult to make transparent connection to the realization of
bang-bang decoupling. To alleviate the need for strong and impulsive
control actions, a dynamical decoupling using only bounded-strength
Hamiltonian is recently proposed.\cite{viola:037901} Within this
framework the same group symmetrization is achieved by exploiting the
{\em Eulerian cycles on a Cayley graph of
  $\mathcal{G}$},\cite{bollobas98} where $\mathcal{G}$ is the
decoupling group. On the other hand a complimentary, geometric
perspective of bang-bang decoupling is recently proposed to provide a
more intuitive picture and provide a method to estimate implementation
errors.\cite{byrd:2002,byrd:012324} In this work we extend the
geometric picture of bang-bang decoupling to the case of continuous
decoupling with bounded controls.  We show that the decoupling by
symmetrization is equivalent to average out error vectors whose
trajectories are determined by the control Hamiltonians. The
decoupling pulses can be intuitively designed by viewing the effect of
control Hamiltonian from a geometric perspective.  To avoid confusing
we will use the term {\em bang-bang decoupling} to refer to the
typical decoupling scheme originally
proposed.\cite{LVEK99,LVSL99,PZ99,LVEK00} The term {\em Eulerian
  decoupling} corresponds to the decoupling with bounded Hamiltonian
introduced by L. Viola and E.  Knill\cite{viola:037901}, while the
term {\em continuous decoupling} corresponds to the version of
continuous decoupling from a geometric picture developed in this work.

The structure of the paper is the following.  In Sec.
\ref{sec:review} we review the original bang-bang decoupling, the
Eulerian decoupling, and the geometric perspective of the bang-bang
decoupling. In Sec. \ref{sec:CDD} we introduce the idea of continuous
dynamical decoupling from a geometric perspective.  In Sec.
\ref{sec:example} several examples are given to illustrate how to
construct the decoupling pulses.  We summarize in Sec.
\ref{sec:summary}.

\section{Review}
\label{sec:review}

\subsection{Bang-bang decoupling decoupling}
In general, dynamical decoupling seeks to eliminate the decoherence of
an open quantum system by effectively averaging out the interaction
between the system and the bath by introducing some strong periodic
control Hamiltonian on the system.\cite{LVEK00,LVEK99,LVSL99,PZ99}
Following the standard treatment, the dynamics of an quantum open
system is determined by the total
Hamiltonian
\begin{equation}
  \label{eqn:H_0}
  H_0=H_S \otimes \Bid_B + \Bid_S \otimes H_B + H_{SB},
\end{equation}
where
\begin{equation}
  H_{SB}=\sum_\gamma S_\gamma \otimes B_\gamma
\end{equation}
represents the system-bath interaction.  Now introduce a control
Hamiltonian $H_c(t)=H_c(t) \otimes \Bid_B$ acting on the system alone.
Denote $U_c(t)$ the time-evolution operator associated with the
control Hamiltonian
\begin{equation}
  U_c(t)\equiv T exp \left\{-i\int_0^t du H_c(u) \right\},
\end{equation}
where $T$ is the time-ordering operator. In periodic dynamical
decoupling one is restricted to the situation where the control field is
{\em cyclic}, i.e., $U_c(t)=U_c(t+T_c)$ for some period $T_c$.  For
any state $|\psi(t)\rangle$ and any operator $O$ in Schr\"odinger picture,
the corresponding state $|\hat{\psi}(t)\rangle$ and operator $\hat{O}(t)$
in the interaction representation associated with $H_c(t)$ are:
\begin{equation}
  |\hat{\psi}(t)\rangle=U^\dagger_c(t)|\psi(t)\rangle
  =U^\dagger_c(t) U(t)|\psi(0)\rangle,
\end{equation}
and
\begin{equation} 
  \hat{O}(t)=U^\dagger_c(t) O U_c(t),
\end{equation}
where $U(t)=T exp\left\{-i H t \right\}$ is the time-evolution operator
associated with $H=H_0+H_c$.
It is easy to show that
\begin{equation}
  |\hat{\psi}(t)\rangle
  =Texp \left\{-i\int_0^t du \hat{H}_0(u) \right\} |\psi(0)\rangle,
\end{equation}
where $\hat{H}_0(t)=U^\dagger_c(t) H_0 U_c(t)$.
Using the periodic condition $U_c(T_c)=\Bid_S$ one finds
\begin{equation}
  |\psi(T_c)\rangle=U(T_c)|\psi(0)\rangle
  =U^\dagger_c(T_c) U(T_c)|\psi(0)\rangle= |\hat{\psi}(T_c)\rangle.
\end{equation}
The {\em stroboscopic} dynamics at $T_N=NT_c$ for integer $N$ 
is hence determined by
the time evolution operator
\begin{equation}
  \hat{U}(T_c)\equiv Texp \left\{-i\int_0^{T_c} du \hat{H}_0(u) \right\}. 
\end{equation}

It is possible to define a k-th order average Hamiltonian $\bar{H}^k$
by applying the Magnus expansion\cite{magnus54} to $\hat{U}(T_c)$, results in
\begin{equation}
  \hat{U}(T_c)=e^{-i \left[\bar{H}^0+\bar{H}^1+\dots \right]T_c}.
\end{equation}
In particular we have
\begin{equation}
  \bar{H}^0=\frac{1}{T_c} \int_0^{T_c} du \hat{H}_0(u)
  =\frac{1}{T_c} \int_0^{T_c} du U^\dagger_c(u) H_0(u) U_c(u),
\end{equation}
and
\begin{equation}
  \bar{H}^1=-\frac{i}{T_c} \int_0^{T_c} dv \int_0^v du 
  \left[\hat{H}_0(v),\hat{H}_0(u)\right].
\end{equation}
Higher-order corrections can also be systematically evaluated.  In the
limit of $T_c \rightarrow 0$, which is the ideal limit of bang-bang
decoupling, $\bar{H}^0$ becomes the dominant term. It can be viewed as
an effective Hamiltonian governing the stroboscopic dynamics under
control Hamiltonian.  In the following we will focus on designing the
effective Hamiltonian $\bar{H}^0$.

Denote $\mathcal{H}_S$ the finite-dimensional state space associated with the system. The idea of decoupling by symmetrization is to identify 
a discrete decoupling group $\mathcal{G}=\{g_j\}$, $j=\{0,\dots,|\mathcal{G}|-1\}$,
acting on $\mathcal{H}_S$ via a faithful, unitary, projective 
representation $\mu$ such that 
$\mu(\mathcal{G})\subset \mathcal{U}(\mathcal{H}_S)$, 
the unitary matrices acting on $\mathcal{H}_S$.
The bang-bang decoupling via $\mathcal{G}$ is implemented by assigning
\begin{equation}
  U_c((l-1)\Delta t+s)=\mu(g_l), s\in [0,\Delta t),
\end{equation}
with $T_c=|\mathcal{G}|\Delta t$, and  $l=\{0,\dots,|\mathcal{G}|-1\}$.
With this assignment the effective Hamiltonian $\bar{H}^0$ becomes:
\begin{equation}
  \bar{H}^0=
  \sum_\gamma \bar{S}_\gamma \otimes B_\gamma,
\end{equation}
where
\begin{equation}
  \bar{S}_\gamma \equiv \Pi_\mathcal{G}(S_\gamma) =
  \frac{1}{|\mathcal{G}|} \sum_{g_j\in \mathcal{G}}
  \mu(g_j)^\dagger S_\gamma \mu(g_j).
\end{equation}
The nontrivial work is to identify the group $\mathcal{G}$ such that
for all $\gamma$ the effective error operator
$\bar{S}_\gamma=\lambda_\gamma \Bid_S$ where $\lambda_\gamma$ is a
real number. Once this is accomplished then the effective Hamiltonian
$\bar{H}^0$ is reduced to
\begin{equation}
  \bar{H}^0= \Bid_S \otimes \sum_\gamma \lambda_\gamma B_\gamma.
\end{equation}
As a result the system is effectively decoupled from the bath, or
equivalently the decoherence is suppressed. Note that in this
formulation the underlying control Hamiltonian is never explicitly
mentioned. An instantaneous, arbitrary strong control Hamiltonian is
needed to implement the desired $U_c(t)$. However, physically it is
impossible to implement such an instantaneous control pulse.  An
arbitrary strong control Hamiltonian would also inevitably induce
transition to higher energy states which are neglected when a two
level approximation is used to describe the qubit space.  Those
unphysical requirements represent some of the main drawbacks of the
original bang-bang decoupling framework.

\subsection{Eulerian decoupling}

In order to alleviate the unphysical requirements of the original
bang-bang decoupling framework, L. Viola and E. Knill propose a general
framework in which the same group symmetrization can be achieved while
using only bounded control Hamiltonians.\cite{viola:037901} Physically
it uses only bounded control Hamiltonians to steer the
time evolution operator. Mathematically it exploits the Eulerian
cycles on a Cayley graph\cite{bollobas98} of the decoupling group
$\mathcal{G}$.  Given a decoupling group $\mathcal{G}$, the first step
in implementing Eulerian decoupling is to find a generating set
$\mathcal{F}=\{f_\alpha \},\alpha=1,\dots, |\mathcal{F}| $, for the
decoupling group $\mathcal{G}$.  The physical implementation
requirement is the ability to generate $f_\alpha$ by some control
Hamiltonian $h_\alpha(t)$ over a period of time $\Delta t$,
\begin{equation}
  f_\alpha=T exp\left \{
    -i\int_0^{\Delta t} du h_\alpha(u)
    \right \}
    , \alpha=1,\dots, |\mathcal{F}|.
\end{equation}

If we image each group element $g_i\in \mathcal{G}$ as a vertex, then
$f_\alpha$ can be imaged as the directional, colored edge connecting
the vertices. If $g^\prime=f_\alpha g$, then we draw a line from point
$g$ to pint $g^\prime$ with color $\alpha$. An Eulerian cycle is
defined as a cycle that uses each edge exactly once. In this case, one
can show that it is always possible to find Eulerian cycle, having
length $L=|\mathcal{G}||\mathcal{F}|$. \cite{bollobas98}
A well-defined Eulerian cycle
beginning at the identity $g_0$ of $\mathcal{G}$ can be uniquely specified
by the sequence of the edge colors used, $P_E=(p_1,p_2,\dots,p_L)$, where
$p_l \in \mathcal{F}$. An Eulerian decoupling is then implemented by 
letting $T_c=L\Delta t$ and by assigning $U_c(t)$ as follows:
\begin{equation}
  U_c(t)[(l-1)\Delta t+s]=u_l(s)U_c[(l-1)\Delta t],
\end{equation}
where $s\in[0,\Delta t)$, and $u_l(s)=Texp\{-i\int_0^s du h_l(u)\}$,
$u_l(\Delta t)=\mu(p_l)$. 
In this way the average Hamiltonian $\bar{H}^0$ becomes:
\begin{equation}
  \bar{H}^0=
  \sum_\gamma \bar{S}_\gamma \otimes B_\gamma,
\end{equation}
where
\begin{eqnarray}
  &\bar{S}_\gamma 
  =& \nonumber  
  \frac{1}{|\mathcal{F}||\mathcal{G}|\Delta t} 
  \int_0^{|\mathcal{F}||\mathcal{G}|\Delta t} du 
  U^\dagger_c(t) H_0(t) U_c(t)  \\
  &=& \nonumber
  \frac{\sum_{g_j\in\mathcal{G}}}{|\mathcal{G}| |\mathcal{F}| \Delta t} 
  \mu(g_j)^\dagger \left\{
  \sum_{\alpha=1}^{|\mathcal{F}|}
  \int_0^{\Delta t} dt u^\dagger_\alpha(t) S_\gamma u_\alpha(t) 
  \right \} \mu(g_j).
\end{eqnarray}

It can be shown that the same decoupling can be achieved through this
average.\cite{viola:037901} Assuming that $\Delta t$ remains the same
as in the Eulerian decoupling, the length of the Eulerian decoupling
is lengthened by a factor of $|\mathcal{F}|$ compared to the bang-bang
decoupling.  Eulerian decoupling provides a guideline to design the
control Hamiltonian if the decoupling group and it's representation is
known. However the search for the decoupling group remains a
nontrivial work.

\subsection{The geometric perspective of the bang-bang decoupling}
Bang-bang decoupling and Eulerian decoupling schemes make heavy use
of the abstract group theory. In recent years a complementary geometric
perspective is developed to provide a more intuitive picture.\cite{byrd:2002}
The geometric picture of the bang-bang decoupling utilities the
Homomorphic mapping between the Lie group $SU(n)$ and $SO(N)$, where
$N=n^2-1$. Let $\lambda_i$, $i=1,\dots ,N$ be the $N$ traceless,
Hermitian generators of $SU(n)$. 
The generators $\{\lambda_j\}$ satisfy trace-orthogonality,
\begin{equation}
  \mathrm{Tr}(\lambda_i \lambda_j)=M\delta_{ij},
\end{equation}
where $M$ is a normalization constant. For any group element $U \in SU(n)$,
one can define a rotation $R \in SO(N)$ via
\begin{equation}
  U^\dagger \lambda_i U=\sum_{j=1}^N R[U]_{ij} \lambda_j.
\end{equation}
This defines a homomorphism from $SU(n)$ to a subgroup of $SO(N)$.

Without loss of generality, one can re-write $H_0$ as
\begin{eqnarray}
  & & H_0 \\
  &=&  \nonumber
  H_S \otimes \Bid_B + \Bid_S \otimes H_B + H_{SB} \\
  &=& \nonumber
  \mathrm{Tr}(H_S) \Bid_S \otimes \Bid_B + (H_S-\mathrm{Tr}(H_S)\Bid)\otimes \Bid_B + \Bid_S \otimes H_B \\
  &+& \nonumber
  \Bid_S \otimes \sum_\gamma \mathrm{Tr}(S_\gamma) B_\gamma
  +\sum_\gamma (S_\gamma-\mathrm{Tr}(S_\gamma)) \otimes B_\gamma \\
  &=& \nonumber
  E_0 +\Bid_s \otimes H^\prime_B+\sum_{\gamma^\prime} S^\prime_\gamma \otimes B^\prime_\gamma,
\end{eqnarray}
where $S^\prime_\gamma$ is traceless. Note that the first term $E_0=\mathrm{Tr}(H_S)$ only
gives rise to an overall phase and can be discarded.
For simplicity (and without loss of generality) we will write 
$H_{SB}=\sum_\gamma S_\gamma \otimes B_\gamma$, where $S_\gamma$ is traceless.

Any traceless system operator $S_\gamma \in SU(n)$ can be expanded in terms of
$\lambda_j$, yielding:
\begin{equation}
  S_\gamma=\sum_i (s_\gamma)_i\lambda_i \equiv \vec{s}_\gamma \cdot \vec{\lambda},
\end{equation}
where
\begin{equation}
  (s_\gamma)_i=\frac{1}{M}\mathrm{Tr}(\lambda_i S_\gamma).
\end{equation}
In other word, a traceless system operator can be represented by a
$N$-dimensional vector .  Using this result the system-bath Hamiltonian
$H_{SB}$ can be written as follows:
\begin{equation}
  H_{SB}
  =\sum_\gamma S_\gamma \otimes B_\gamma
  =\sum_\gamma \left( \vec{s}_\gamma \cdot \vec{\lambda} \right)\otimes B_\gamma,
\end{equation}
where $\vec{s}_\gamma$ is a vector of length $N$.  In the following we
should refer $\vec{s}_\gamma$ as error vectors.  By using error
vectors to represent the system-bath Hamiltonian $H_{SB}$, the
decoupling by symmetrization over a group $\mathcal{G}$ with projective
representation $\mu(g_i)$ can be viewed as an average over rotated 
error vectors. Following this line the average Hamiltonian
$\bar{H}^0$ becomes
\begin{eqnarray}
  \bar{H}^0
  &=&
  \sum_\gamma \frac{1}{|\mathcal{G}|} 
  \sum_{g\in \mathcal{G}}
  \mu(g)^\dagger S_\gamma \mu(g) \otimes B_\gamma \\
  &=& \nonumber
  \sum_\gamma
  \frac{1}{|\mathcal{G}|} 
  \sum_{g\in \mathcal{G}}
  \mu(g)^\dagger 
  \sum_i (\vec{s}_\gamma)_i \lambda_i
  \mu(g) \otimes B_\gamma \\
  &=& \nonumber
  \sum_\gamma
  \frac{1}{|\mathcal{G}|} 
  \sum_{g\in \mathcal{G}}
  \sum_{ij} (\vec{s}_\gamma)_i R[\mu(g)]_{ij} \lambda_j \otimes B_\gamma \\
  &=& \nonumber
  \sum_\gamma
  \sum_j \left\{
  \frac{1}{|\mathcal{G}|} 
  \sum_{g\in \mathcal{G}}
  \sum_i R^\dagger[\mu(g)]_{ji} (\vec{s}_\gamma)_i 
  \right \} \lambda_j \otimes B_\gamma \\
  &=& \nonumber
  \sum_\gamma 
  \left( \vec{s^\prime}_\gamma \cdot \vec{\lambda} \right)
  \otimes B_\gamma,
\end{eqnarray}
where the average error vector $\vec{s^\prime}_\gamma$ is equal to
\begin{equation}
  \vec{s^\prime}_\gamma
  = 
  \frac{1}{|\mathcal{G}|} 
  \sum_{g\in \mathcal{G}}
  \sum_i R^\dagger[\mu(g)]_{ji} (\vec{s}_\gamma)_i 
  = 
  \frac{1}{|\mathcal{G}|} 
  \sum_{g\in \mathcal{G}}
  R^\dagger[\mu(g)] \vec{s}_\gamma .
\end{equation}
From the geometric perspective the decoupling condition (in case
of quantum memory) is equal to require that the average error
$\vec{s}_\gamma$ be zero for all $\gamma$. Geometrically each term in
group symmetrization procedure corresponds to an effective rotation
$R^\dagger[\mu(g)]\in O(N)$ on all error vectors.  However it is
evident that the error vector can be averaged to zero by a set of
rotations which do not correspond to the representation of some
decoupling group. It is thus intriguing to discuss if a underlying
group structure is necessary to achieve dynamical decoupling.

\section{Continuous decoupling from a geometric perspective}
\label{sec:CDD}
The geometric picture of the bang-bang decoupling is intuitive but it
shares the same drawback as the bang-bang decoupling, i.e., the error
vector is instantaneously rotated to another vector by some rotation.
This drawback can be alleviated if a bounded control Hamiltonian is
used to continuously rotate the error vector. We thus seek to
formulate a framework for continuous dynamical decoupling from a
geometric perspective. Recall that the average Hamiltonian has
the following expression
\begin{equation}
\bar{H}^0=\frac{1}{T_c} \int_0^{T_c} du U^\dagger_c(u) H_0(u) U_c(u)
=\sum_\gamma \bar{S}_\gamma \otimes B_\gamma.
\end{equation}
Now instead of piecewisely mapping $U_c(t)$ into the representation of
some decoupling group, one represents the effective system operator
$\bar{S}_\gamma$ by its corresponding average error vector
$\vec{s^\prime}_\gamma$. The average error vector can be expressed as
the time average over the trajectory of the error vector rotated by 
$R^\dagger[U_c]$:
\begin{eqnarray}
  \bar{S}_\gamma 
  &=& 
  \frac{1}{T_c} 
  \int_0^{T_c} du U^\dagger_c(u) S_\gamma U_c(u)  \\
  &=& \nonumber
  \frac{1}{T_c} 
  \int_0^{T_c} du  \sum_i (s_\gamma)_i U^\dagger_c(u) \lambda_i U_c(u)  \\ 
  &=& \nonumber
  \frac{1}{T_c} 
  \int_0^{T_c} du  \sum_{ij} (s_\gamma)_i R[U_c(u)]_{ij} \lambda_j \\
  &=& \nonumber
  \sum_j \left\{
    \frac{1}{T_c} 
    \int_0^{T_c} du \sum_i  R^\dagger[U_c(u)]_{ji}(s_\gamma)_i  
  \right\} \lambda_j \\
  &=& \nonumber
  \vec{s^\prime}_\gamma \cdot \vec{\lambda},
\end{eqnarray}
where 
\begin{equation}
  (s^\prime_\gamma)_j=
  \frac{1}{T_c} \int_0^{T_c} du \sum_i  R^\dagger[U_c(u)]_{ji}(s_\gamma)_i . 
\end{equation}
Or using vector notation:
\begin{equation}
  \vec{s^\prime}_\gamma 
  =\frac{1}{T_c} \int_0^{T_c} du R^\dagger[U_c(u)] \vec{s}_\gamma. 
\end{equation}

The decoupling condition (in case of quantum memory) is to require
$\vec{s^\prime}_\gamma=0$ for all $\gamma$. Note that we don't
explicitly require that $U_c$ be the representation of some group.  In
order to make the design of control Hamiltonian easier it is desirable
to make a more transparent connection between the decoupling and the
control Hamiltonian.  First define a time-dependent error vector via
\begin{equation}
  \vec{s}\cdot U^\dagger_c(t)\vec{\lambda}U_c(t)
  \equiv\vec{s}(t)\cdot\vec{\lambda}.
\end{equation}
The $i$-th component of $\vec{s}(t)$ can be expressed as
\begin{equation}
  s_i(t)=\frac{1}{M}\mathrm{Tr}\left\{
      \lambda_i\vec{s}\cdot U^\dagger_c(t)\vec{\lambda}U_c(t)
    \right\}.
\end{equation}
It is instructive to study the trajectory of $\vec{s}(t)$ when the
control Hamiltonian is proportional to one of the generators of
$SU(n)$. Assuming that $H_l(t)=a_l (t) \lambda_l$, where $a_l(t)$ 
represents the envelope function of the control pulse, one finds
\begin{eqnarray}
  \frac{d }{dt} Ms_i(t)
  &=& 
  \mathrm{Tr}\left\{ \lambda_i
      ia_l(t)\lambda_l \vec{s}\cdot U^\dagger_c(t)\vec{\lambda}U_c(t)\right\} \\
  &+& \nonumber
  \mathrm{Tr}
  \left\{ \lambda_i
    \vec{s}\cdot U^\dagger_c(t)\vec{\lambda}U_c(t)(-ia_l(t))\lambda_l) 
  \right\} \\
  &=& \nonumber
  i a_l(t) \mathrm{Tr}\left\{ \lambda_i
    \left[ \lambda_l, \sum_j s_j(t)\lambda_j \right] \right\} \\
  &=& \nonumber
  i a_l(t) \sum_j s_j(t) \mathrm{Tr}
  \left\{ \lambda_i i\sum_k f_{ljk}\lambda_k \right\} \\
  &=& \nonumber
  i a_l(t) \sum_j i f_{lji} s_j(t)
  \equiv
  -i a_l(t) \left[L_l\right]_{ij} s_j(t),
\end{eqnarray}
where we have defined
\begin{eqnarray}
  \left[L_l\right]_{ij}=-if_{lji}=+if_{lij},
\end{eqnarray}
and $f_{ijk}$ is the structure function of $SU(n)$. From the $O(N)$
point of view, the effect of control Hamiltonian $a_l(t)\lambda_l$ is
to induce a rotation with generator $L_l$ and with speed $a_l(t)$. It
is thus useful to  express $L_i$ in terms of the natural generators of
$SO(N)$. The natural generator of $SO(N)$ are antisymmetric Hermitian
matrices $L_{\mu \nu}$ where $\mu,\nu=1, \cdots N$, whose components
have the form
\begin{equation}
  [L_{\mu \nu}]_{ij}=-i(\delta_{\mu i}\delta_{\nu j}-\delta_{\mu j}\delta_{\nu i}).
\end{equation}
We will restrict ourself to $\mu <\nu$ as a convention and to avoid
double counting. They satisfy the commutation relation
\begin{eqnarray}
& & [L_{\mu \nu}, L_{\mu^\prime \nu^\prime}] \\
&=& \nonumber
-i\left(
   \delta_{\mu \nu^\prime} L_{\nu \mu^\prime}
  -\delta_{\mu \mu^\prime} L_{\nu \nu^\prime}
  +\delta_{\nu \nu^\prime} L_{\mu \mu^\prime}
  -\delta_{\nu \mu^\prime} L_{\mu \nu^\prime}
\right),
\end{eqnarray}
and the trace orthogonality condition
\begin{equation}
\mathrm{Tr}(L_{\mu \nu} L_{\mu^\prime \nu^\prime})
=M\delta_{\mu \mu^\prime} \delta_{\nu \nu^\prime}.
\end{equation}

Giving a control Hamiltonian of the form $a_l(t) \lambda_l$, in the
geometric picture it corresponds to a time independent rotation
generator $L_l$ and an time-dependent
envelope function $a_l(t)$ representing the
time dependent speed of the rotation. By expressing $L_l$ in terms
of the natural generators of $SO(N)$
\begin{equation}
  L_l=\sum_{\mu \nu} X_{\mu \nu} L_{\mu \nu},
\end{equation}
a compact notation of the form
\begin{equation}
  [\lambda_l,\vec{\lambda}]=
  \left(\sum_{\mu \nu} X_{\mu \nu} L_{\mu \nu}\right)\vec{\lambda}
\end{equation}
can be used to represent the effect of control Hamiltonian $\lambda_l$.
For example, for $SU(3)\rightarrow SO(8)$ one has
\begin{equation}
  [\lambda_3,\vec{\lambda}]=
  \left(-L_{12}+L_{45}+L_{76}\right)\vec{\lambda}.
\end{equation}
It immediately leads us to the conclusion that $\lambda_3$ can be
used to average any vector in the 1-2 plan to zero with appropriate
envelope function. In the appendix we explicitly calculate the
$[\lambda_l,\vec{\lambda}]$ for $SU(2),SU(3)$, and $SU(4)$.
Using the continuous decoupling from a geometric perspective,
the design of the decoupling pulse sequences can be outlined
as follows: The first step is to find the corresponding error
vectors from the system-bath Hamiltonian. The second step is
to identify the useful control Hamiltonian $\lambda_l$. The third step
is to design a proper envelope function $a(t)$ to ensure the
average error vector to be zero.

Continuous decoupling is robust against implementation imperfection
because a small implementation error on the control Hamiltonian will
only result in a small deviation of the average error vectors from
their ideal values. The implementation error can be evaluated via the
distance between the ideal average error vector and the real average
error vectors. Let $\vec{s^\prime}_{\lambda,id}$ be ideal average
error vectors resulted from a perfect control Hamiltonian while
$\vec{s^\prime}_{\lambda}$ be the real average error vectors resulted
from a imperfect control Hamiltonian. The Euclidean distance 
between two vectors 
$d(\vec{s^\prime}_{\lambda,id},\vec{s^\prime}_{\lambda})
\equiv \sqrt{
(\vec{s^\prime}_{\lambda,id}-\vec{s^\prime}_{\lambda}) \cdot
(\vec{s^\prime}_{\lambda,id}-\vec{s^\prime}_{\lambda})
}$
can be used to quantify the implementation error.



\section{Examples}
\label{sec:example}

In this section we study several examples which illustrate
the basic idea of continuous decoupling from geometric perspective.

\subsection{Single qubit spin-flip decoherence}
Consider a single qubit with a single error operator,
$\{S_\alpha\}=\{\sigma_z\}$. In geometric picture it corresponds to a
single error vector $\vec{s}=(0,0,1)$. Intuitively, performing any
$2\pi$ rotation in a plan containing the error vector $\vec{s}$ should average
the error vector to zero. Using the results in the appendix it is easy
to verify that one can choose the control Hamiltonian to be
proportional to $\lambda_1(=\sigma_x)$ or $\lambda_2(=\sigma_y)$. From the
geometric perspective, this corresponds to rotate the error vector in
$x$-$z$ or $y$-$z$ plan using $SO(3)$ generator $L_{13}$ or $L_{23}$.
If we were to choose $\sigma_y$ and assume that $H_c(t)=a(t)\sigma_y$,
the time-dependent error vector becomes
\begin{equation}
\vec{s}(t)=e^{-i\int_0^t du a(u) L_{13}} \vec{s}(0)=e^{-iA(t)L_{13}}\vec{s}(0),
\end{equation}
where $A(t)=\int_0^t du a(u)$. The decoupling condition can be written as
\begin{equation}
\frac{1}{T_c}\int_0^{T_c} du e^{-iA(t)L_{13}} \vec{s}(0)=\vec{0}.
\end{equation}
The decoupling condition can be satisfied very generally 
by requiring that
\begin{equation}
e^{-iA(T_c/2)L_{13}}=R[\pi,\hat{y}],  
\end{equation}
and
\begin{equation}
  A(T_c/2+t)=\pi+A(t), t=[0,T_c/2),
\end{equation}
where $R[\pi,\hat{y}]$ represents a $\pi$ rotation around $y$-axis.

Geometrically this corresponds to rotate the error vector from
$\vec{s}$ to $-\vec{s}$ at some speed controlled by $a(t)$, and rotate
it back to $\vec{s}$ with the same speed profile.  The first condition
ensures that the error vector is steered to $-\vec{s}$ at half time
$t=T_c/2$. The second condition ensures that the contribution from the
second half cancels exactly the contribution from the first half,
resulting in zero average error vector.

It is known that the minimal decoupling group of bang-bang decoupling
in this case is the
group $Z_2=\{e,g\}$ where $g^2=e$. It is also referred as parity kick
in the literature. The corresponding representation can be chosen to be
either $\{\mu(e)=\Bid,\mu(g)=\sigma_x\}$ or
$\{\mu(e)=\Bid,\mu(g)=\sigma_y\}$. From the geometric picture the
effect of $\mu(g)$ is to rotate error vector $\vec{s}$ instantaneously to
$-\vec{s}$, ensuring the average error vector be zero.  It is easy to
show that a larger group $C_4=\{e,g^2,g^3,g^3\}$, where $g^4=e$, can
achieve the same decoupling.  The corresponding representation is
$\{\mu(e)=\Bid,\mu(g)=r,\mu(g^2)=r^2,\mu(g^2)=r^3\}$ where $r$ is the
$\pi/2$ rotation along x-direction. From the geometric picture this
corresponds to rotate the error vector by $\pi/2$ at each kick.  From
the typical bang-bang decoupling point of view larger decoupling
group represents a less optimal decoupling scheme since more kicks are
needed. On the other hand when viewed as the limiting case of the continuous decoupling,
$Z_2$ bang-bang decoupling corresponds to require a large amplitude $a(t)$
for small $t$ and turn $a(t)$ off once the error vector $\vec{s}$ is
steered to $-\vec{s}$. A different $a(t)$ can be similarly designed to
reproduce the $C_4$ bang-bang decoupling. However in any real 
physical implementation there is a upper limit for the 
strength of $a(t)$. Hence there is a minimal time needed to finish 
one continuous decoupling cycle and the
ideal limit of bang-bang decoupling is never reached. 
It is this
minimal $T_c$, not the order of the decoupling group, when compared to
the decoherence time, indicates the efficiency of the decoupling
scheme. 

\subsection{Single qubit full decoherence}
\begin{figure}
  \includegraphics[scale=0.35]{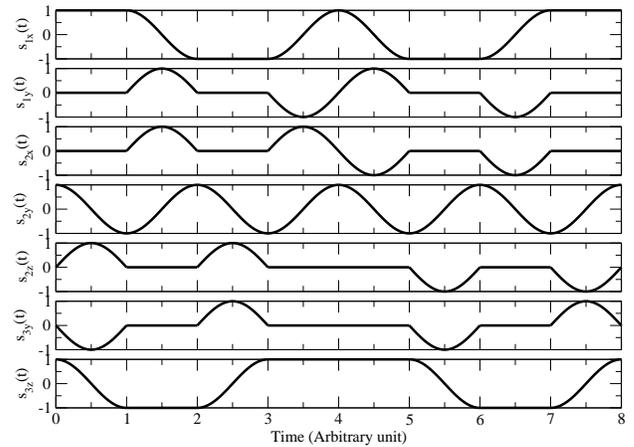}
  \caption{Trajectories of error vectors under the decoupling sequence described
in example B. It is assumed that the envelope function in each sub-period
is the same.
\label{fig:single qubit}}
\end{figure}

Consider next a single qubit with all possible error operators,
$\{S_\alpha\}=\{\sigma_x, \sigma_y, \sigma_z\}$.  In geometric picture
they correspond to three error vectors $\vec{s}_1=(1,0,0)$,
$\vec{s}_2=(0,1,0)$, and $\vec{s}_3=(0,0,1)$.  Intuitively a $2\pi$
rotation in a plan containing both the vectors $\vec{s}_1$ and
$\vec{s}_2$ can average both error vectors to zero, but $\vec{s}_3$
will remain unchanged. However it is possible to design a sequence
which average all three error vectors to zero. To see how such a pulse
sequence can be constructed, first recall that in previous example the
vector $\vec{s}_i(t)$ must reach $-\vec{s}_i$ during the decoupling
operation. We thus seek to rotate $\vec{s}_i$ to $-\vec{s}_i$ using
alternating generators, in hope that the residual errors will cancel
each other when a full decoupling cycle is finished. Using this idea
it is straightforward to construct and verify that a sequence of 
$\pi$ rotations using following generators
$\left\{\lambda_1,\lambda_3,\lambda_1,\lambda_3,
  \lambda_3,\lambda_1,\lambda_3,\lambda_1\right\}$ can average all
three error vectors to zero. In Fig.\ref{fig:single qubit} we plot the
trajectories of three error vectors during the decoupling sequence.  It
is evident from the figure that a judicial envelope function will
ensure that all time integrals of error vectors are zero.

The decoupling pulse sequence designed here is equivalent to the
Eulerian decoupling prescribed in Ref \onlinecite{viola:037901}.
However as pointed out in previous example the efficiency
consideration should be based on the continuous decoupling framework.
In Ref \onlinecite{byrd:2002}, an example is given to demonstrate that
a decoupling group may not be necessary to achieve the decoupling
condition. From the geometric picture of the bang-bang decoupling it
suffices to rotate instantaneously the error vectors to the vortices
of a tetrahedron. From continuous decoupling point of view it is
evident that for this kind of decoupling sequences the residual errors
will accumulate when the error vectors are rotated from one vertex to
another. It is difficult to cancel these residuals systematically
using a design similar to what has been done in this example. Even if
it is accomplished, it would represent a less optimal solution. We
thus argue that a decoupling group is not necessary to achieve
continuous dynamical decoupling but a decoupling group usually
provides a guideline to design the optimal continuous decoupling
sequence.

\subsection{Two qubits independent decoherence}
Next example consists of a two qubit system with independent dephasing
error operators. The system-bath Hamiltonian has the form:
\begin{equation}
  H_{SB}=g_1 \left(\sigma_z \otimes \Bid)\right)\otimes B_1
  +g_2 \left(\Bid\otimes\sigma_z\right)\otimes B_2.
\end{equation}
The Hamiltonian corresponds to two error vectors
$\vec{s}_1=g_1 \vec{\lambda}_3$ and $\vec{s}_2=g_2 \vec{\lambda}_6$.
Any rotation in $SO(15)$ which can rotate both error vectors
by $2\pi$ should effectively decouple the system from the
dephasing error. In many solid-state qubit, Heisenberg
exchange is utilized to implement quantum operations. It is thus
advantageous to design a decoupling sequence which is compatible
with Heisenberg exchange interaction. In terms of Pauli matrices
the Heisenberg interaction is written as
\begin{equation}
  H_{ex}=J \sigma_x\otimes\sigma_x
  +J \sigma_y\otimes\sigma_y+J \sigma_z\otimes\sigma_z.
\end{equation}
In other words the Heisenberg interaction corresponds to three vectors $\vec{\lambda}_7$,
$\vec{\lambda}_{11}$, and $\vec{\lambda}_{15}$ from the geometric
perspective. A decoupling
sequence compatible with Heisenberg interaction should
average error vectors to zero while leave those three
vectors in tact. Using the results in the appendix it is
apparent that one should avoid using the generators
$\lambda_1 \cdots \lambda_6$, $\lambda_8 \cdots \lambda_9$,
and $\lambda_{12} \cdots \lambda_{14}$ which either can't average
error vectors to zero or have undesirable side effects on Heisenberg
interaction. This leaves us to the only choice
of $\lambda_7$, $\lambda_{11}$, and $\lambda_{15}$.
It is then easy to verify that any one of those three generators can
be used as the control Hamiltonian to average out the dephasing
error with an appropriately designed envelope function $a(t)$ while
leaves the Heisenberg interaction in tact.

\subsection{One qubit with one ancilla level}
Our final example consists of a three level system in which
the first two levels $|1\rangle, |2\rangle$ are used as the qubit space 
while the third level $|3\rangle$ is used as an ancilla level used to implement quantum operation in qubit space. 
We assume that all three levels are coupled to the environment,
with the following system-bath Hamiltonian:
\begin{eqnarray}
  H_{SB}&=&g_1 \left( |1\rangle\langle 1|-|2\rangle\langle 2| \right)\otimes B_1\\
  &+& \nonumber
  g_2 \left( |1\rangle\langle 3|+|3\rangle\langle 1| \right)\otimes B_2
  + g_3 \left( |2\rangle\langle 3|+|3\rangle\langle 2| \right)\otimes B_3.
\end{eqnarray}
This system-bath Hamiltonian corresponds to three error vectors
$\vec{s}_1=\vec{\lambda_3}$, $\vec{s}_2=\vec{\lambda_4}$, 
and $\vec{s}_3=\vec{\lambda_6}$. By consulting the $SU(2)\rightarrow SO(8)$
mapping in the appendix it is easy to show that all three error vectors 
can be averaged to zero via the generator 
$\lambda_2$. This simple example illustrates that the continuous dynamical
decoupling from geometric perspective can be straightforwardly applied not only
to the logical qubit space but also the physical qubit space or the larger space
in which the physical qubit space is embedded in.

\section{Summary and discussion}
\label{sec:summary}
In summary we have developed a continuous dynamical decoupling
framework from a geometric perspective. Within this framework, the
need to perform unphysical, arbitrarily strong and fast control is
eliminated, and bang-bang control can be viewed as a unphysical limit
of the continuous decoupling in which the control Hamiltonian is
unbounded. The decoupling condition is equal to require that all
time-dependent error vectors $\vec{s}_\gamma$ average to zero at the
end of the decoupling cycle.  The trajectories of the error vectors
are steered by the control Hamiltonian. By expressing control
Hamiltonian in terms of the generators of $SU(n)$, the geometric
picture provides an intuitive way to design the decoupling pulse
sequences, provided that the structure function of the corresponding
$SU(n)$ is known.  Several examples are given to explicitly illustrate
how to design the decoupling sequences starting from a given
system-bath Hamiltonian.  We also show that, in stead of the order of
the decoupling group, the minimal time needed to finish one continuous
decoupling cycle.  The framework is not restricted to physical qubit
space or logical qubit space.  The ancilla levels or other relevant
nearby levels can be naturally included in the analysis. This is
important especially when ancilla levels are needed to perform quantum
operation or when the nearby levels can not be neglected.

The paper has focused on designing the decoupling pulses which are
ideal for quantum memory. Less is addressed about how to perform
quantum operation and decoupling at the same time. The continuous
decoupling framework developed in this work can be extended to systematically
treat this problem.\cite{byrd:012324}. The detail of the
analysis is beyond the scope of this work and will be presented
elsewhere. It is also intriguing to discuss the issue of optimal
control. In a typical bang-bang decoupling framework a decoupling
group is identified, assuming that the corresponding time evolution
operators can be implemented.  However in real system the possible
control Hamiltonian at our disposal might be limited and the desired
time evolution operators might not be achievable. It is thus nature to
ask the complementary questions: Giving a set of possible control
Hamiltonian, what is the optimal decoupling sequence? The geometric
continuous decoupling framework developed here is suitable to answer
this question. As a simplest example consider the case where some of
the generators of $SU(n)$ can not be used as the control Hamiltonian.
In this case one can use the remaining generators and their
corresponding rotations to construct a (sub)-optimal decoupling
sequence.

Recently there has been efforts to unify the dynamical decoupling and
the quantum Zeno effect.\cite{facchi:032314} Roughly speaking
bang-bang control and quantum Zero effect both require a strong
interaction with a quantum system. This idea shared by two scheme
leads to the possible unification. It is thus very interesting to see
if the continuous decoupling scheme in which the arbitrarily strong
pulses are not necessary can still be connected to the quantum Zeno
effect.

\appendix*

\section{$SU(n)$ and $SO(N)$}

In this appendix we explicitly calculate the correspondence between
the generators of $SU(n)$ and the natural generators of $SO(N)$ where
$N=n^2-1$ using the procedure outlined in Sec \ref{sec:CDD}. We 
adapt the following convention for the natural generators of $SO(N)$:
\begin{equation}
  [L_{\mu \nu}]_{ij}=-i(\delta_{\mu i}\delta_{\nu j}-\delta_{\mu j}\delta_{\nu i}),
\end{equation}
where $\mu <\nu$. Note that the choise of the generators of $SU(n)$ is not 
unique. Unitary transformation on a set of generators results in another
set of generators. However this only amounts to an rotation on error vectors
and the basis vectors of $SO(N)$.

\subsection{$SU(2)$ and $SO(3)$}
\label{sec:su2}
Let $\lambda_1=\sigma_x$, $\lambda_2=\sigma_y$, and $\lambda_3=\sigma_z$ 
be the 3 generators of $SU(2)$. They satisfy the trace-orthogonality
\begin{equation}
  \mathrm{Tr}(\lambda_i \lambda_j)=2 \delta{ij}.
\end{equation}
It is easy to verify that 
\begin{eqnarray}
&& [\lambda_1, \vec{\lambda}]=+4L_{23} \vec{\lambda}, \\
&& [\lambda_2, \vec{\lambda}]=-4L_{13} \vec{\lambda}, \\
&& [\lambda_3, \vec{\lambda}]=+4L_{12} \vec{\lambda}.
\end{eqnarray}

\subsection{$SU(3)$ and $SO(8)$}
A standard set of generators of the $SU(3)$ are the Gell-Mann matrices:
\begin{equation} 
\lambda_1=
\left( \begin{array}{ccc}
0 & 1 & 0 \\
1 & 0 & 0 \\
0 & 0 & 0
\end{array} \right),
\lambda_2=
\left( \begin{array}{ccc}
0 & -i & 0 \\
i & 0 & 0 \\
0 & 0 & 0
\end{array} \right),
\lambda_3=
\left( \begin{array}{ccc}
1 & 0 & 0 \\
0 & -1 & 0 \\
0 & 0 & 0
\end{array} \right),
\end{equation}

\begin{equation}
\lambda_4=
\left( \begin{array}{ccc}
0 & 0 & 1 \\
0 & 0 & 0 \\
1 & 0 & 0
\end{array} \right)
\lambda_5=
\left( \begin{array}{ccc}
0 & 0 & -i \\
0 & 0 & 0 \\
i & 0 & 0
\end{array} \right),
\lambda_6=
\left( \begin{array}{ccc}
0 & 0 & 0 \\
0 & 0 & 1 \\
0 & 1 & 0
\end{array} \right),
\end{equation}

\begin{equation}
\lambda_7=
\left( \begin{array}{ccc}
0 & 0 & 0 \\
0 & 0 & -i \\
0 & i & 0
\end{array} \right),
\lambda_8=\frac{1}{\sqrt{3}}
\left( \begin{array}{ccc}
1 & 0 & 0 \\
0 & 1 & 0 \\
0 & 0 & -2
\end{array} \right).
\end{equation}
 They satisfy the trace-orthogonality
\begin{equation}
  \mathrm{Tr}(\lambda_i \lambda_j)=2 \delta{ij}.
\end{equation}
Using this set of generators one can verify the following relations
\begin{eqnarray}
&&[\lambda_1,\vec{\lambda}]=(+4L_{23}+2L_{47}+2L_{65}) \vec{\lambda}, \\ \nonumber
&&[\lambda_2,\vec{\lambda}]=(-4L_{13}+2L_{46}+2L_{57}) \vec{\lambda}, \\ \nonumber
&&[\lambda_3,\vec{\lambda}]=(+4L_{21}+2L_{45}-2L_{76}) \vec{\lambda}, \\ \nonumber
&&[\lambda_4,\vec{\lambda}]=(-2L_{62}-2L_{71}-2L_{53}-2\sqrt{3}L_{58}) \vec{\lambda}, \\ \nonumber
&&[\lambda_5,\vec{\lambda}]=(+2L_{16}+2L_{34}-2L_{72}-2\sqrt{3}L_{84}) \vec{\lambda}, \\ \nonumber
&&[\lambda_6,\vec{\lambda}]=(-2L_{24}+2L_{37}+2L_{51}+2\sqrt{3}L_{78}) \vec{\lambda}, \\ \nonumber
&&[\lambda_7,\vec{\lambda}]=(+2L_{14}+2L_{25}-2L_{63}-2\sqrt{3}L_{86}) \vec{\lambda}, \\ \nonumber
&&[\lambda_8,\vec{\lambda}]=(+2\sqrt{3}L_{45}+2\sqrt{3}L_{67}) \vec{\lambda}.
\end{eqnarray}

\subsection{$SU(4)$ and $SO(15)$}
For $SU(n)$ where $n$ is a power of 2, it is convenient to use the product of 
Pauli matrices to form the generator of $SU(n)$. We hence use the following
assignment for the generators of $SU(4)$:

\begin{eqnarray}
&& 
\lambda_1 =\sigma_1 \otimes \Bid, 
\lambda_2 =\sigma_2 \otimes \Bid,  
\lambda_3 =\sigma_3 \otimes \Bid, \\
&& \nonumber
\lambda_4 =\Bid \otimes \sigma_1,
\lambda_5 =\Bid \otimes \sigma_2,
\lambda_6 =\Bid \otimes \sigma_3, \\
&& \nonumber
\lambda_7 =\sigma_1 \otimes \sigma_1,
\lambda_8 =\sigma_1 \otimes \sigma_2,
\lambda_9 =\sigma_1 \otimes \sigma_3, \\
&& \nonumber
\lambda_{10} =\sigma_2 \otimes \sigma_1,
\lambda_{11} =\sigma_2 \otimes \sigma_2,
\lambda_{12} =\sigma_2 \otimes \sigma_3, \\
&& \nonumber
\lambda_{13} =\sigma_3 \otimes \sigma_1,
\lambda_{14} =\sigma_3 \otimes \sigma_2,
\lambda_{15} =\sigma_3 \otimes \sigma_3. \\
\end{eqnarray}

The generators satisfy the trace-orthogonality
\begin{equation}
\mathrm{Tr}(\lambda_i \lambda_j)=4\delta_{ij}.
\end{equation}

By matrix manipulation one can verify that the following
relations holds
\begin{eqnarray*}
& & [\lambda_1,\vec{\lambda}]=
\left( 
+8L_{2,3}+8L_{10,13}+8L_{11,14}+8L_{12,15}
\right)\vec{\lambda} \\
& & [\lambda_2,\vec{\lambda}]=
\left( 
-8L_{1,3}-8L_{7,13}-8L_{8,14}-8L_{9,15}
\right)\vec{\lambda} \\
& & [\lambda_3,\vec{\lambda}]=
\left( 
+8L_{1,2}+8L_{7,10}+8L_{8,11}+8L_{9,12}
\right)\vec{\lambda} \\
& & [\lambda_4,\vec{\lambda}]=
\left( 
+8L_{5,6}+8L_{8,9}+8L_{11,12}+8L_{14,15}
\right)\vec{\lambda} \\
& & [\lambda_5,\vec{\lambda}]=
\left( 
-8L_{4,6}-8L_{7,9}-8L_{10,12}-8L_{13,15}
\right)\vec{\lambda} \\
& & [\lambda_6,\vec{\lambda}]=
\left( 
+8L_{4,5}+8L_{7,8}+8L_{10,11}+8L_{13,14}
\right)\vec{\lambda} \\
& & [\lambda_7,\vec{\lambda}]=
\left( 
+8L_{2,13}-8L_{3,10}+8L_{5,9}-8L_{6,8}
\right)\vec{\lambda} \\
& & [\lambda_8,\vec{\lambda}]=
\left( 
-8L_{2,14}-8L_{3,11}-8L_{4,9}+8L_{6,7}
\right)\vec{\lambda} \\
& & [\lambda_9,\vec{\lambda}]=
\left( 
+8L_{2,15}-8L_{3,12}+8L_{4,8}-8L_{5,7}
\right)\vec{\lambda} \\
& & [\lambda_{10},\vec{\lambda}]=
\left( 
-8L_{1,13}+8L_{3,7}+8L_{5,12}-8L_{6,11}
\right)\vec{\lambda} \\
& & [\lambda_{11},\vec{\lambda}]=
\left( 
-8L_{1,14}+8L_{3,8}-8L_{4,12}+8L_{6,10}
\right)\vec{\lambda} \\
& & [\lambda_{12},\vec{\lambda}]=
\left( 
-8L_{1,15}+8L_{3,9}+8L_{4,11}-8L_{5,10}
\right)\vec{\lambda} \\
& & [\lambda_{13},\vec{\lambda}]=
\left( 
+8L_{1,10}-8L_{2,7}+8L_{5,15}-8L_{6,14}
\right)\vec{\lambda} \\
& & [\lambda_{14},\vec{\lambda}]=
\left( 
+8L_{1,11}-8L_{2,8}-8L_{4,15}+8L_{6,13}
\right)\vec{\lambda} \\
& & [\lambda_{15},\vec{\lambda}]=
\left( 
+8L_{1,12}-8L_{2,9}+8L_{4,14}-8L_{5,13}
\right)\vec{\lambda} \\
\end{eqnarray*}

\begin{acknowledgments}
  We acknowledge the support of National Science Council in Taiwan through
  grant NSC 93-2112-M-007-038.
\end{acknowledgments}


\end{document}